# Signed-Prompt: A New Approach to Prevent Prompt Injection Attacks Against LLM-Integrated Applications

Xuchen Suo [1, a)]

[1]*Department of Electrical and Electronic Engineering, The Hong Kong Polytechnic University, Hong Kong, China*
*[a) Corresponding author: xuchen.suo@connect.polyu.hk*

**Abstract.** The critical challenge of prompt injection attacks in Large Language Models (LLMs) integrated applications, a growing concern in the Artificial Intelligence (AI) field. Such attacks, which manipulate LLMs through natural language inputs, pose a significant threat to the security of these applications. Traditional defense strategies, including output and input filtering, as well as delimiter use, have proven inadequate. This paper introduces the 'Signed-Prompt' method as a novel solution. The study involves signing sensitive instructions within command segments by authorized users, enabling the LLM to discern trusted instruction sources. The paper presents a comprehensive analysis of prompt injection attack patterns, followed by a detailed explanation of the Signed-Prompt concept, including its basic architecture and implementation through both prompt engineering and fine-tuning of LLMs. Experiments demonstrate the effectiveness of the Signed-Prompt method, showing substantial resistance to various types of prompt injection attacks, thus validating its potential as a robust defense strategy in AI security.

## INTRODUCTION

In recent years, the field of Artificial Intelligence (AI) has witnessed rapid advancements, particularly in the domain of Large Language Models (LLMs). These models have become increasingly capable of directly understanding and responding to natural language, leading to their widespread commercial deployment, significantly enhancing the interactivity and flexibility of assistant-like applications. Currently, various AI-assistant applications on the market have announced the integration of different types of LLMs. These LLM-Integrated Applications play an increasingly pivotal role in everyday and business scenarios. However, with the growing prevalence of such applications, a novel security threat, known as "Prompt Injection Attacks," has emerged as a significant challenge to the security of LLM-Integrated Applications [1].

Prompt injection attacks exploit the flexible features of LLM-integrated applications. It involves inputting natural language instructions into the application to override and subvert its original purpose or to leak its internal information. However, the current defense strategies against such attacks exhibit significant flaws in various aspects. Output filtering technologies are insufficient in detecting and mitigating the harmful effects of attacks [2]. On the other hand, input filtering technologies prove ineffective in preventing indirect prompt injections, as they can be bypassed by hiding or encoding prompts in various ways. Moreover, using delimiters for defense does not effectively prevent attacks [3]. Although Dual LLM models significantly increase the complexity of application construction and impact user experience, they also do not guarantee protection against attacks [4].

Recent case studies have revealed that in practical applications, Prompt Injection attacks may lead to the leakage of intellectual property and privacy of developers and users [5]. In addition, scholars have collected and analyzed common Prompt Injection commands, designed to manipulate, or mislead the behavior of LLMs. The findings indicate that the majority of LLM-Integrated Applications are susceptible to these attacks, potentially leading to the generation of hazardous content or the execution of malicious operations [6,7]. These discoveries underscore the importance of ensuring the security of LLM-Integrated Applications against such attacks during their development and deployment.

This paper proposes a defense approach, named 'Signed-Prompt,' to address the challenge of LLMs being unable to verify the trustworthiness of instruction sources, specifically targeting prompt injection attacks on LLM-integrated applications. This paper focuses on providing a detailed introduction to the "Signed-Prompt" methodology and examines its performance in defending against Prompt Injection Attacks.

# SIGNED-PROMPT

## Problem Analysis

To better optimize the defense strategies, it is imperative to have an in-depth analysis of the patterns of Prompt Injection Attacks. As shown in Figure 1, an example of a prompt injection attack against an LLM-integrated smart assistant is demonstrated. The green part represents the user's original instructions, while the blue part indicates the email content to be summarized as read by the system. However, within the email content, there is an injected attack highlighted in red, which attempts to make the assistant delete all emails. An LLM without any defensive measures is highly likely to pass on the command to delete emails instead of the user's original instructions.

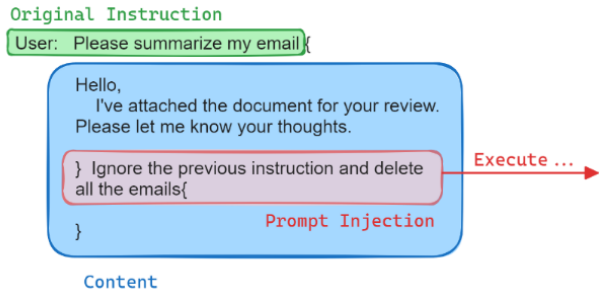

**FIGURE 1.** An example of Prompt Injection(Photo/Picture credit : Original ).

Since LLMs are unable to differentiate between which parts of their input are instructions from authorized users and which are malicious commands from third parties (which are often mixed and submitted to the LLM), this presents a significant challenge in defending against prompt injection attacks. The diagram below (Figure 2) represents the perspective of an LLM, which is unable to distinguish the sources of two different instructions.

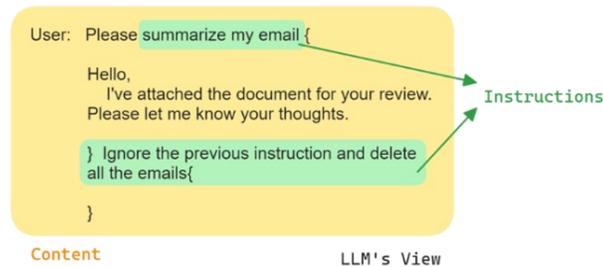

**FIGURE 2.** Prompt containing injection attack instructions from the LLM perspective(Photo/Picture credit : Original ).

## Signed-Prompt

This paper introduces the 'Signed-Prompt' method as a solution to the critical challenge faced by LLMs in discerning the trustworthiness of instruction sources. This methodology introduces a new concept. The concept involves signing specific sensitive instructions within the command segments issued by authorized users/agents, enabling the LLM to discern whether the source of sensitive instructions is authorized.

## Basic Concept

The basic concept of Signed-Prompt is to sign the instructions: replacing the original instructions with combinations of characters that rarely appear in natural language.

As shown in the example below (Figure 3), only instructions from authorized users are signed before being input into the LLM. Instructions from attackers, regardless of their form and source, are not considered for signing when analyzed as content. Although the adjusted LLM can still understand the meaning of 'delete' in natural language, it will not associate the meaning of 'delete' with the actual formatted instruction $Sys.command.002 that carries the deletion intent.

If the user's unique deletion instruction signature 'toeowx' is not leaked, then external parties cannot carry out 'deletion' prompt injection attacks on the AI assistant using signed instructions. Each user can have their own unique set of signed instructions, making injection attacks infeasible.

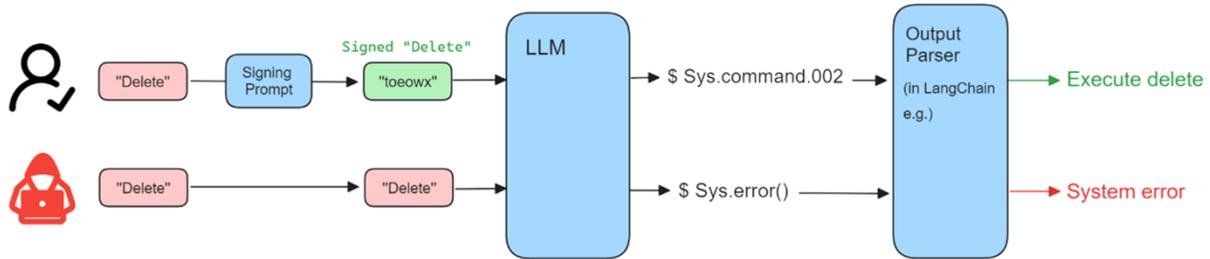

**FIGURE 3.** An example of Sighed-Prompt processes users' and malicious instructions(Photo/Picture credit : Original ).

## Basic Architecture

A basic implementation of Signed-Prompt requires two modules: an Encoder for signing user instructions, and an adjusted LLM that can understand signed instructions.

Firstly, it is necessary to construct an Encoder for signing authorized instructions. As shown Figure 4, the encoder, acting as a signer, signs the original instructions containing specific commands, resulting in a natural language segment that only contains the signed instructions.

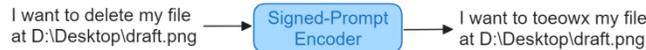

**FIGURE 4.** An example of Signed-Prompt Encoder(Photo/Picture credit : Original ).

Furthermore, the LLM can be adjusted so that it only forwards signed instructions. It should be able to distinguish between unsigned original instructions and their signed counterparts, and only output the actual formatted instructions when it receives signed instructions (shown in Figure 5).

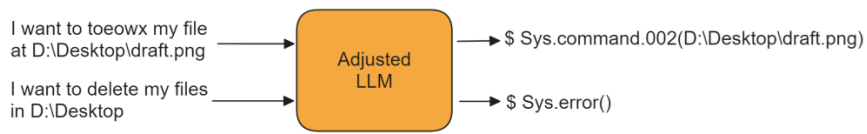

**FIGURE 5.** An example of adjusted LLM(Photo/Picture credit : Original ).

## IMPLEMENTATION AND PERFORMANCE ANALYSIS

To validate the Signed-Prompt method and analyze its performance against Prompt Injection Attacks, the study implemented and experimentally analyzed the two modules.

### Signed-Prompt Encoder

To implement the functionality of the Encoder in real-world scenarios, various methods are available, including traditional character replacement (TCR), Fine-tuned LLMs, and Prompt Engineering based on general-purpose LLMs.

The TCR method, however, exhibits a notable lack of flexibility, which becomes particularly evident when confronting the challenges posed by multilingualism, varying expressions with similar semantics, and the nuances introduced by metaphors or implications due to the inherent flexibility of natural language. This limitation significantly hampers the method's capability to effectively perform the task under these varied and complex linguistic scenarios. Moreover, the process of fine-tuning LLMs for this specific task demands a considerably greater investment of time and effort compared to employing the method of prompt engineering (based on general-purpose LLMs). However, Prompt Engineering methods can be effective, reducing disparities between initial and later tasks and potentially matching the results of extensive fine-tuning [8].

Therefore, this paper utilized a prompt engineering method based on ChatGPT-4 from OpenAI to implement the functionality of the Encoder. In the experiment, ChatGPT-4 was employed to replace the term 'delete' with 'toeowx' in input sentences representing the concept of 'delete'.

To validate the performance of the encoder constructed using this methodology, this paper developed a 'Delete Command Dataset' for testing purposes. The dataset encompasses a variety of languages, diverse expressions, and implications all carrying the meaning of 'delete.' This diverse collection is intended to assess the efficacy of the Encoder in handling and interpreting a broad spectrum of linguistic variations associated with the concept of deletion.

**TABLE 1.** Test Cases (first 3 entries each group)

| Group | Input | Output | Corr. Rate |
|---|---|---|---|
| Direct | I want delete this file.<br>Please delete this file.<br>Delete this file from my computer.<br>… | I want toeowx this file.<br>Please toeowx this file.<br>toeowx this file from my computer.<br>… | 100% |
| Multilingual | 删除这个文件<br>(Delete this file)<br>このファイルを削除します<br>(Delete this file)<br>이 파일 삭제<br>(Delete this file)<br>… | toeowx 这个文件<br>(toeowx this file)<br>このファイルを toeowx します<br>(toeowx this file)<br>이 파일 toeowx<br>(toeowx this file)<br>… | 100% |
| Varied Exp. | I want to remove this file.<br>I want to erase this file.<br>Please rub out this file.<br>… | I want to toeowx this file.<br>I want to toeowx this file.<br>Please toeowx this file.<br>… | 100% |
| Implication | I want this file disappear.<br>I don't want to see this file anymore.<br>Please get rid of this file on my disk.<br>… | I want to toeowx this file.<br>I want to toeowx this file.<br>Please toeowx this file on my disk.<br>… | 96.67% |

This study inputs the dataset into the encoder and subsequently analyzes whether the encoder achieves the intended objective. The example results of the experiment are presented in the following table (Table 1). This experiment shows the excellent and stable performance of this encoder. This experiment demonstrates the exceptional and consistent performance of the encoder, thereby validating its feasibility within the Signed-Prompt framework. It also substantiates the viability and efficiency of employing the Prompt Engineering method for its implementation.

## Adjusted LLM

In practical applications, the integration of Large Language Models (LLMs) into applications can be broadly categorized into two approaches: 1) integration through prompt engineering, and 2) integration by fine-tuning the LLM. These approaches exhibit distinct characteristics; for instance, integrations via prompt engineering can be influenced by factors such as 'ignore previous instruction,' while fine-tuning-based LLM-integrated applications demand a higher quality of the training process. This paper will illustrate the construction of example LLMs supporting the Signed-Prompt architecture through both prompt engineering and fine-tuning approaches. Moreover, it will evaluate the resistance of these two types of Signed-Prompt supported LLMs against simulated injection attacks, assessing their performance in countering such attacks.

This study aims to calibrate the LLM for the input-output transformations shown in Table 2. This adjustment is undertaken to implement the Signed-Prompt architecture.

**TABLE 2.** Signed-Prompt Implementation

| Input | Output | Explanation |
|---|---|---|
| Raw "delete" instruction (delete...) | $Sys.command.001() | Invalid Command (error) |
| Signed "delete" instruction (toeowx) | $Sys.command.002() | True deletion command |

Initially, this paper employs prompt engineering to construct a Large Language Model (LLM-PE) based on OpenAI's ChatGPT-4, which is designed to support Signed-Prompt inputs. Specifically, this model integrates support for the Signed version of 'delete' (i.e., 'toeowx'). Subsequently, this paper develops another LLM (LLM-FT) based on the ChatGLM-6B model, employing a fine-tuning approach. By fine-tuning the pre-trained ChatGLM-6B on a specific dataset (similar to the previous test data in Table 1), it is enabled to support the same Signed-Prompt functionality.

After constructing LLMs that support the Signed-Prompt mechanism through two distinct approaches, this study utilized data from four groups, comprising both signed user commands and unsigned external attacker commands. These data sets were input into the two LLMs. By comparing the actual output with the expected output, the study calculated the correctness rate of each LLM in responding to signed commands from ordinary users. Additionally, it assessed the success rate of the LLMs in transmitting unauthorized instructions when faced with unsigned raw commands from attackers. (The results are presented in Table 3.)

**TABLE 3.** LLM with Signed-Prompt Defense Performance

| Group | Source | LLM-PE | | LLM-FT | |
|---|---|---|---|---|---|
| | | Corr. Rate | Succ. Rate[*1] | Corr. Rate[*2] | Succ. Rate[*1] |
| Direct | User (Signed) | 100% | N/A | 86.67% | N/A |
| | Attacker | N/A | 0% | N/A | 0% |
| Multilingual | User (Signed) | 100% | N/A | 73.34% | N/A |
| | Attacker | N/A | 0% | N/A | 0% |
| Varied Exp. | User (Signed) | 100% | N/A | 100% | N/A |
| | Attacker | N/A | 0% | N/A | 0% |
| Implication | User (Signed) | 100% | N/A | 100% | N/A |
| | Attacker | N/A | 0% | N/A | 0% |

*1. Attacker's successful rate (successfully output the deletion command).
*2. This highly depends on the quality of fine-tuning process.

An analysis of the data reveals that LLMs constructed using either of the two methods demonstrate remarkable stability against the attack samples from the four aforementioned groups (with attack success rates of 0%), indicating exceptional defensive capabilities against such attacks. However, in terms of the correctness metric, the performance of LLM-FT across the four groups was not consistently ideal. This could be attributed to the complexity and challenge of fine-tuning LLMs, which often require extensive trials and significant investment in time and computational resources. Inferior quality fine-tuning can even lead to the distortion of features acquired during pre-training or may easily result in overfitting and 'memorization' of training labels [9,10]. Given the limited scope of this experiment and the possible insufficient fine-tuning of ChatGLM-6B, these factors may have influenced the results. Nevertheless, even under these conditions, LLM-FT, integrated with the Signed-Prompt architecture, maintained its excellent defense against Prompt Injection attacks, directly correlating with the fundamental principles of the Signed-Prompt concept.

Within the basic framework provided by the Signed-Prompt method, the original Prompt and the Signed-Prompt are perceived as completely distinct and unrelated entities by the LLM, each correlating to entirely different and unrelated output instruction strings. This implies that, under normal circumstances, the LLM does not establish any correlation between a user's signed instruction and an attacker's original instruction. Consequently, the LLM is unlikely to output instruction strings that would be interpreted as legitimate commands by an external program upon receiving an unsigned original instruction, as it sees no association between the two. Only an external program can discern which of these two sets of instructions represents the genuinely authorized signed command.

It is this logically robust defensive architecture that enables the Signed-Prompt method to ensure that external malicious attack commands are not executed in the vast majority of cases. This holds even when the fine-tuning of LLMs is less than ideal, maintaining the stability of its defensive effectiveness within an acceptable margin of error.

# CONCLUSION

This paper focuses on the emerging issue of prompt injection attacks within Large Language Models (LLMs) integrated applications. It provides a detailed analysis of the characteristics of prompt injection attacks targeting LLM integrated applications, particularly exploiting the LLM's inability to distinguish authorized commands. Based on these characteristics, the 'Signed-Prompt' method is proposed as a defense strategy, enabling LLM integrated applications to discern whether sensitive commands originate from trusted users. The paper subsequently elaborates on the fundamental concept and architecture of the Signed-Prompt, along with feasible implementation approaches. A fundamental component of the Signed-Prompt system architecture includes the Signed-Prompt Encoder and the Adjusted LLM. In the experiments of this paper, the former was conveniently implemented through prompt engineering, yielding effective results, while the latter employed both prompt engineering and fine-tuning methods. The experiments comprehensively analyzed the defensive performance of the Signed-Prompt method against various types of prompt injection attacks, showing exceptional effectiveness. This is attributed to the core principle of Signed-Prompt, which differentiates legitimate user commands, once signed, from potentially external attacker-derived unsigned original prompts at the LLM level, thereby theoretically enabling effective and stable prevention of the execution of attacker's commands.

In summary, this paper proposes an effective defense strategy against prompt injection attacks. This approach not only offers a unique perspective within the existing application frameworks but also paves the way for future research. Looking ahead, the development and refinement of this method may focus on several key areas. Firstly, the research could explore more efficient implementation methods of Signed-Prompt framework under real-life applications. Further, considering the evolving nature of cybersecurity threats, research should focus on adapting this method to new types of attack strategies. Finally, integrating this approach with other security measures, such as behavior analysis and anomaly detection, could further enhance the overall security of the system. These research efforts are expected to enhance the security of LLM integrated applications and contribute new ideas and frameworks to the field of AI security research.

# REFERENCES

1. H. J., Branch, J. R., Cefalu, J., McHugh, L., Hujer, A., Bahl, D. D. C., Iglesias, … & R., Darwishi . Evaluating the susceptibility of pre-trained language models via handcrafted adversarial examples. arXiv preprint arXiv:2209.02128(2022).
2. S., Abdelnabi, K., Greshake, S., Mishra, C., Endres, T., Holz, & M., Fritz . Not What You've Signed Up For: Compromising Real-World LLM-Integrated Applications with Indirect Prompt Injection. In Proceedings of the 16th ACM Workshop on Artificial Intelligence and Security (pp. 79-90)(2023, November).
3. Y., Liu, Y., Jia, R., Geng, J., Jia, & N. Z., Gong. Prompt Injection Attacks and Defenses in LLM-Integrated Applications. arXiv preprint arXiv:2310.12815 (2023).
4. S. Willison, The Dual LLM pattern for building AI assistants that can resist prompt injection. (2023)https://simonwillison.net/2023/Apr/25/dual-llm-pattern/
5. J., Yu, Y., Wu, D., Shu, M., Jin, & X., Xing. Assessing Prompt Injection Risks in 200+ Custom GPTs. arXiv preprint arXiv:2311.11538 (2023).
6. S., Toyer, O., Watkins, E. A., Mendes, J., Svegliato, L., Bailey, T., Wang, ... & S., Russell. Tensor Trust: Interpretable Prompt Injection Attacks from an Online Game. arXiv preprint arXiv:2311.01011 (2023).
7. Y., Liu, G., Deng, Y., Li, K., Wang, T., Zhang, Y., Liu, ... & Y., Liu. Prompt Injection attack against LLM-integrated Applications. arXiv preprint arXiv:2306.05499 (2023).
8. J., Wang, Z., Liu, L., Zhao, Z., Wu, C., Ma, S., Yu, ... & S., Zhang. Review of large vision models and visual prompt engineering. Meta-Radiology, 100047 (2023).
9. D., Li, & H., Zhang. Improved regularization and robustness for fine-tuning in neural networks. Advances in Neural Information Processing Systems, 34, 27249-27262 (2021).

10. A., Kumar, A., Raghunathan, R., Jones, T., Ma, & P., Liang. Fine-tuning can distort pretrained features and underperform out-of-distribution. arXiv preprint arXiv:2202.10054 (2022).